\documentclass[12pt]{article}
\usepackage{epsfig}
\usepackage{amssymb}

\begin{document}

\newcommand{\beq}{\begin{equation}}
\newcommand{\eeq}{\end{equation}}
\newcommand{\barr}{\begin{eqnarray}}
\newcommand{\earr}{\end{eqnarray}}

\newcommand{\andy}[1]{ }

\newcommand{\ket}[1]{| #1 \rangle}
\newcommand{\bra}[1]{\langle #1 |}
\def\As{{\cal A}}
\def\cZ{{\cal Z}}
\renewcommand{\Re}{{\rm Re}}
\renewcommand{\Im}{{\rm Im}}

\begin{titlepage}

\vspace{.5cm}
\begin{center}
{\LARGE Quantum Zeno phenomena: pulsed versus continuous
measurement}

\quad

{\large P. Facchi and S. Pascazio
\\
           \quad \\
Dipartimento di Fisica, Universit\`a di Bari \\
and Istituto Nazionale di Fisica Nucleare, Sezione di Bari \\
 I-70126 Bari, Italy \\
}

\vspace*{.5cm}

{\small\bf Abstract}\\ \end{center}

{\small The time evolution of an unstable quantum mechanical
system coupled with an external measuring agent is investigated.
According to the features of the interaction Hamiltonian, a
quantum Zeno effect (hindered decay) or an inverse quantum Zeno
effect (accelerated decay) can take place, depending on the
response time of the apparatus. The transition between the two
regimes is analyzed for both pulsed and continuous measurements. }

\vspace*{.5cm} PACS: 03.65.Xp

\end{titlepage}

\newpage

\section{Quantum Zeno effect: fundamentals}
\label{sec-dpw}
\andy{sec-dpw}

Let $H$ be the total Hamiltonian of a quantum system. The survival
probability of the system in state $\ket{a}$ is
\andy{uno}
\beq
P(t) = |\As (t)|^2 =|\langle a|e^{-iHt}|a\rangle |^2.
\label{eq:uno}
\eeq
An elementary expansion yields a quadratic behavior at short times
\andy{quadratic}
\beq
P(t) \sim 1 - t^2/\tau_{\rm Z}^2, \qquad \tau_{\rm Z}^{-2}
\equiv \langle a|H^2|a\rangle - \langle a|H|a\rangle^2 ,
\label{eq:quadratic}
\eeq
where $\tau_{\rm Z}$ is called Zeno time. Observe that if one
divides the Hamiltonian into a free and an interaction part $H=H_0
+ H_{\rm I}$, with $H_0\ket{a}=\omega_a\ket{a}$ and $\bra{a}H_{\rm
I}\ket{a}=0$, the Zeno time reads $\tau_{\rm Z}^{-2} =
\bra{a}H_{\rm I}^2\ket{a}$ and depends only on the off-diagonal
part of the Hamiltonian.

\begin{figure}[t]
\begin{center}
\epsfig{file=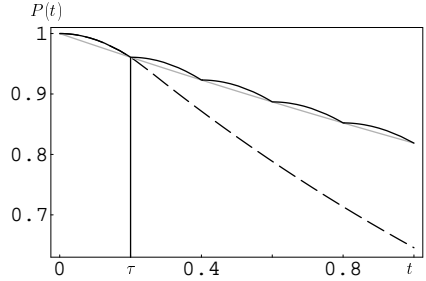,width=5.5cm}
\end{center}
\caption{Evolution with frequent ``pulsed" measurements: quantum
Zeno effect. The dashed (full) line is the survival probability
without (with) measurements. The gray line is the interpolating
exponential (\ref{eq:survN0}).}
\label{fig:zenoevol}
\end{figure}

We first consider ``pulsed" measurements, as in the seminal
approach \cite{Misra}. The complementary notion of ``continuous
measurement" will be discussed in Sec.\
\ref{sec-QZEcont}. Perform $N$ (instantaneous) measurements at
time intervals $\tau=t/N$ (pulsed observation), in order to check
whether the system is still in its initial state $\ket{a}$. The
survival probability after the measurements reads
\andy{survN}
\beq
P^{(N)}(t)=P(\tau)^N = P\left(t/N\right)^N
\sim\exp\left(-t^2/\tau_{\rm Z}^2 N\right)
\stackrel{N \rightarrow\infty}{\longrightarrow} 1 .
\eeq
The (mathematical) limit is the quantum Zeno paradox: ``A watched
pot never boils". For large (but finite) $N$ the evolution is
slowed down (quantum Zeno effect). Indeed, the survival
probability after $N$ pulsed measurements ($t=N\tau$) is
interpolated by an exponential law \cite{heraclitus}
\andy{survN0}
\beq
P^{(N)}(t)=P(\tau)^N=\exp(N\log P(\tau))= \exp(-\gamma_{\rm
eff}(\tau) t) ,
\label{eq:survN0}
\eeq
with an {\em effective decay rate}
\beq
\gamma_{\rm eff}(\tau) \equiv -\frac{1}{\tau}\log P(\tau) \ge0 \;
.
\label{eq:gammaeffdef}
\eeq
For $\tau\to 0 $ one gets  $P(\tau) \sim \exp (-\tau^2/\tau_{\rm
Z}^2)$, whence
\beq
\gamma_{\rm eff}(\tau) \sim \tau/\tau_{\rm Z}^2: \qquad
(\tau\to 0)
\label{eq:lingammaeff}
\eeq
increasingly frequent measurements hinder the evolution and tend
to ``freeze" it. The Zeno evolution is represented in Figure
\ref{fig:zenoevol}.

\section{Unstable systems}
\label{sec-unstable}
\andy{sec-unstable}

Consider the spontaneous decay of state $\ket{a}$ into state
$\ket{b}$ described by the Hamiltonian
\barr
H=H_0+H_{\rm I}=\omega_a\ket{a}\bra{a}+\sum_k
\omega_k a^\dagger_k a_k+\sum_k \phi_k \left( a_k \ket{a}\bra{b}+a^\dagger_k
\ket{b}\bra{a}\right),
\label{eq:hamtot}
\earr
with $\bra{a}a\rangle=\bra{b}b\rangle=1$ and
$[a_k,a^\dagger_{k'}]=\delta_{k k'}$, other commutators = 0. As is
well known, the Fourier-Laplace transform of the survival
amplitude $\As(t)$ in (\ref{eq:uno}) is the expectation value of
the resolvent
\andy{transf}
\beq
\label{eq:transf}
G_a(E)=\int_0^{\infty}dt\;e^{iEt}\As(t)
=\bra{a}\frac{i}{E-H}\ket{a}, \qquad
\As(t)=\int_{\rm B}\frac{dE}{2\pi}\;e^{-iEt}G_a(E) ,
\eeq
the Bromwich path B being a horizontal line $\Im E=$constant$>0$
in the half plane of convergence of the Fourier-Laplace transform
(upper half plane). By performing Dyson's resummation, the
resolvent $G_a$ can be expressed in terms of the self-energy
function $\Sigma_a$
\andy{propag}
\beq
\label{eq:propag}
G_a(E)=\frac{i}{E-\omega_a-\Sigma_a(E)}, \qquad
\Sigma_a(E)=\sum_k
\frac{\left|\phi_k\right|^2}{E-\omega_k}=\int_0^\infty
d\omega\;\frac{\kappa_a(\omega)}{E-\omega}  ,
\eeq
where $\kappa_a(\omega)=\bra{a}H_{\rm I}\delta(\omega-H_0)H_{\rm
I}\ket{a}=\sum_k \left|\phi_k\right|^2\delta(\omega-\omega_k)$ is
the form factor of the interaction (spectral density function).

If $-\Sigma_a(0)<\omega_a$ (which happens for sufficiently smooth
form factors and small coupling), the resolvent is analytic in the
whole complex plane cut along the positive real axis (continuous
spectrum of $H$). On the other hand, there exists a pole $E_{\rm
pole}$ located just below the branch cut in the second Riemann
sheet, solution of the equation $E_{\rm
pole}-\omega_a-\Sigma_{a{\rm II}}(E_{\rm pole})=0$, $\Sigma_{a{\rm
II}}$ being the determination of the self-energy function in the
second sheet. The pole has a real and imaginary part $E_{\rm
pole}=\omega_a + \delta\omega_a-i\gamma/2$ given by
\beq
\delta\omega_a=\Re \Sigma_{a{\rm II}}(E_{\rm pole})\simeq \Re
\Sigma_a(\omega_a+i0^+)={\cal P}\int d\omega \frac{\kappa_a(\omega)}{\omega_a-\omega}
={\cal P}\sum_{k} \frac{|\phi_k|^2}{\omega_a-\omega_k} ,
\label{eq:2shift}
\eeq
\beq
\gamma=-2\Im \Sigma_{a{\rm II}}(E_{\rm pole})\simeq -2\Im
\Sigma_a(\omega_a+i0^+)=2\pi\kappa_a(\omega_a) ,
\label{eq:FGR}
\eeq
up to fourth order in the coupling constant. One recognizes the
second-order energy shift $\delta\omega_a$ and the celebrated
Fermi ``golden" rule $\gamma$ \cite{Fermi}. The survival amplitude
has the general form
\beq
\As(t)=\As_{\rm pole}(t)+\As_{\rm cut}(t),
\eeq
where $\As_{\rm pole}(t)=e^{-i(\omega_a+\delta\omega_a)t-\gamma
t/2}/[1-\Sigma'_{a{\rm II}}(E_{\rm pole})]$, $\As_{\rm cut}$ being
the branch-cut contribution. At intermediate times, the pole
contribution dominates the evolution (Weisskopf-Wigner
approximation \cite{Gamow28}) and
\andy{Psurv}
\beq
P(t)\simeq |\As_{\rm pole}(t)|^2 =  \cZ e^{-\gamma t} ,\qquad
\cZ=\left|1-\Sigma'_{a{\rm II}}(E_{\rm
pole})\right|^{-2} ,
\label{eq:psurv}
\eeq
where $\cZ$, the intersection of the asymptotic exponential with
the $t=0$ axis, is the wave function renormalization. As is well
known the exponential law is corrected by the cut contribution,
which is responsible for a quadratic behavior at short times and a
power law at long times.

\section{Inverse quantum Zeno effect}
\label{sec-IZE}
\andy{sec-IZE}

Consider an unstable system with decay rate $\gamma$ given by
(\ref{eq:FGR}). By performing a single measurement at a
sufficiently long time $t$, when the exponential behavior
$P(t)\simeq e^{-\gamma t}$ is dominant, one infers from
(\ref{eq:survN0}) that the effective decay rate is simply the
natural (undisturbed) one
\beq
\gamma_{\rm eff}(\tau) \stackrel{{\rm ``long"}
\tau}{\longrightarrow} \gamma.
\label{eq:longtau}
\eeq
We now ask whether it is possible to find a {\em finite} time
$\tau^*$ such that
\andy{tstardef}
\beq
\gamma_{\rm eff}(\tau^*)=\gamma.
\label{eq:tstardef}
\eeq
\begin{figure}[t]
\begin{center}
\epsfig{file=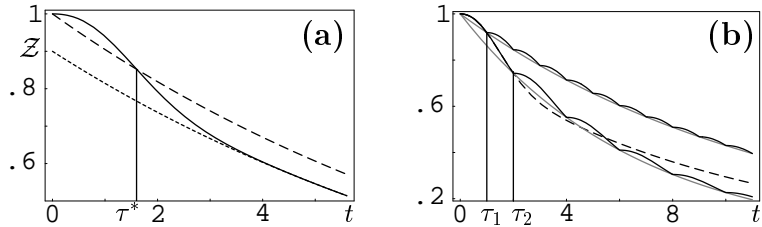,width=10cm}
\end{center}
\caption{(a) Determination of the transition time $\tau^*$.
The full line is the survival probability $P(t)$, the dashed line
the exponential $e^{-\gamma t}$ and the dotted line the asymptotic
exponential $\cZ e^{-\gamma t}$ in (\ref{eq:psurv}). (b) Quantum
Zeno vs inverse Zeno (``Heraclitus") effect. The dashed line
represents the undisturbed survival probability $P(t)$. The full
lines represent the survival probabilities with measurements at
time intervals $\tau$ and the dotted lines their exponential
interpolations (\ref{eq:survN0}). For $\tau_1<\tau^*<\tau_2$ the
effective decay rate $\gamma_{\rm eff}(\tau_1)$ [$\gamma_{\rm
eff}(\tau_2)$] is smaller (QZE) [larger (IZE)] than the ``natural"
decay rate $\gamma$. When $\tau=\tau^*$ one recovers the natural
lifetime, according to (\ref{eq:tstardef}).}
\label{fig:gtau}
\end{figure}
If such a time exists, then by performing measurements at time
intervals $\tau^*$ the system decays according to its undisturbed
decay rate $\gamma$, as if no measurements were performed. The
related concept of ``jump" time was considered in
\cite{Schulman98}. By (\ref{eq:gammaeffdef}) and
(\ref{eq:tstardef}) we get $P(\tau^*)=e^{-\gamma \tau^*}$: the
time $\tau^*$ is the intersection between  the curves $P(t)$ and
$e^{-\gamma t}$. In the situation depicted in Figure
\ref{fig:gtau}(a) such a time $\tau^*$ exists: the full line is
the survival probability $P(t)$ and the dashed line the
exponential $e^{-\gamma t}$ [the dotted line is the asymptotic
exponential $\cZ e^{-\gamma t}$, see (\ref{eq:psurv})]. By looking
at Figure
\ref{fig:gtau}(b) we realize that $\tau^*$ represents a {\it
transition time} from a quantum Zeno to an inverse quantum Zeno
regime \cite{heraclitus}. Indeed
\barr
\mbox{if}\;\;  \tau =\tau_1<\tau^* \quad &\Rightarrow& \quad
\gamma_{\rm eff}(\tau_1)<\gamma\qquad
\mbox{Quantum Zeno Effect (QZE);} \nonumber\\
\mbox{if}\;\;  \tau = \tau_2>\tau^* \quad &\Rightarrow& \quad
\gamma_{\rm eff}(\tau_2)>\gamma\qquad
 \mbox{Inverse quantum Zeno Effect (IZE)} .
 \nonumber
\earr
If $\tau^*$ exists, frequent measurements first accelerate decay
(IZE)
\cite{Kofman,heraclitus}, then, eventually, slow it down (QZE)
when the frequency of measurements becomes larger than $1/\tau^*$
\cite{heraclitus,Raizen01}. Note that the existence of such a transition
time $\tau^*$ is related to the value of the wave function
renormalization $\cZ$: if $\cZ<1$ a finite $\tau^*$ certainly
exists \cite{heraclitus} and the system exhibits both QZE and IZE,
depending on the frequency of measurements. (This is the case
considered in Figure
\ref{fig:gtau}.)
The transition from a Zeno to an inverse Zeno regime has been
recently confirmed in a beautiful experiment performed by Raizen's
group \cite{Raizen01}.

\section{Pulsed versus continuous observation}
\label{sec-QZEcont}
\andy{sec-QZEcont}

We now introduce some alternative descriptions of a measurement
process and discuss the notion of continuous measurement. This is
to be contrasted with the idea of pulsed measurements, discussed
in the previous sections and hinging upon von Neumann's
projections. We will show that the use of instantaneous pulsed
measurements is not essential to obtain QZE
\cite{Kraus81,Sudbery} or (possibly) IZE . We will provide a
dynamical picture of the measurement process by introducing a
Hamiltonian description of the interaction with the detector and
show that the detector response time plays a role very similar to
that of the period between measurements in the pulsed version
\cite{Schulman98}. We will also show that irreversibility is not
an essential ingredient of this picture. By replacing an
irreversible detector with an oscillating one, we show that QZE
and IZE are a simple consequence of a strong interaction between
the ``observed" decaying system and an ``observing" agent (the
``detector") which closely ``looks" at the system
\cite{zenoreview}.

\subsection{Pulsed observation (period $\tau$)}
\label{sec-pulsed}
\andy{sec-pulsed}

\begin{figure}[t]
\begin{center}
\epsfig{file=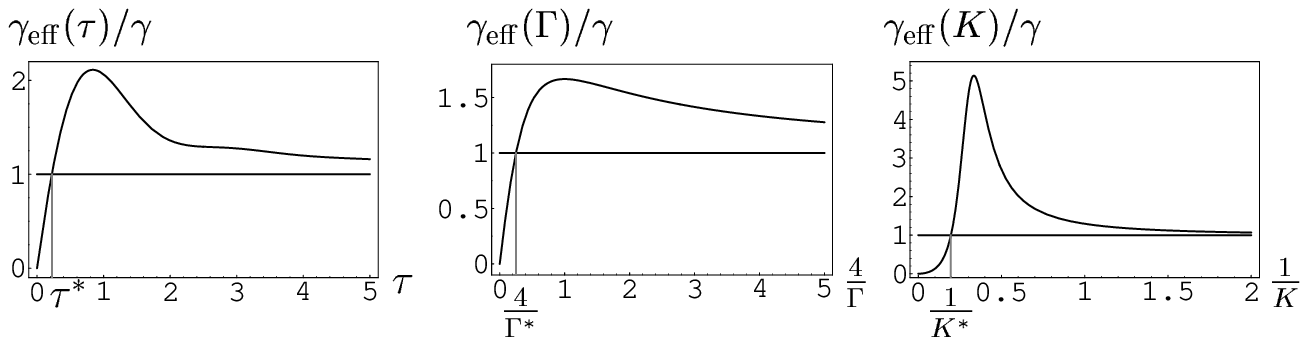,width=\textwidth}
\end{center}
\caption{Effective decay rate as a function of the detector response time: pulsed observation (period
$\tau$); continuous observation (decay time $\Gamma^{-1}$);
continuous Rabi observation (Rabi period $2\pi/K$).}
\label{fig:gammatot}
\end{figure}

We start by considering pulsed measurements performed at time
intervals $\tau$. For simplicity we choose a Lorentzian form
factor
$\kappa_a(\omega)=\lambda^2\Lambda/\pi(\omega^{2}+\Lambda^2)$,
from which an analytical expression of the survival amplitude can
be easily obtained. (Notice that the Hamiltonian in this case is
not lower bounded and one expects no deviations from exponential
behavior at very large times.) We chose $\lambda=0.1$, $\Lambda=1$
and $\omega_a=3$, so that $\cZ=0.998<1$, a finite $\tau^*$ exists
and the system exhibits a QZE-IZE transition. The effective decay
rate (\ref{eq:gammaeffdef}) is shown in the left frame of Figure
\ref{fig:gammatot} as a function of $\tau$.
Notice the linear behavior (\ref{eq:lingammaeff}) for $\tau\to 0$,
with slope $1/\tau_{\rm Z}^2$. Observe that for the chosen value
of the parameters, the linear approximation (\ref{eq:lingammaeff})
is valid well beyond the intersection $\tau^*$ and one gets
$\tau^*\simeq\tau_{\rm Z}^2
\gamma=0.2$. For $\tau>\tau^*$ the system decays faster, with a decay rate
$\gamma_{\rm eff}$ that first increases up to $2 \gamma$, then
decreases and eventually relaxes to the natural decay rate
$\gamma$ according to (\ref{eq:longtau}).

\subsection{Continuous observation (response time $\Gamma^{-1}$)}
\label{sec-cont}
\andy{sec-cont}

Let us consider now a continuous measurement process. This is
accomplished, for instance, by adding to (\ref{eq:hamtot}) the
following interaction Hamiltonian
\beq
H_{\rm meas}(\Gamma)=\sqrt{\frac{\Gamma}{2\pi}}\int d\omega \left(
d(\omega) \ket{b}\bra{M}+d^\dagger(\omega)
\ket{M}\bra{a}\right) ,
\label{eq:intGamma}
\eeq
with $[d(\omega),d^\dagger(\omega')]=\delta(\omega-\omega')$,
other commutators = 0. As soon as it becomes populated, state
$\ket{b}$ decays into state $\ket{M}$, with a decay rate $\Gamma$.
This yields a continuous monitoring of the decay process $\ket{a}
\to \ket{b}$,
with a response time $1/\Gamma$. The presence of the interaction
Hamiltonian (\ref{eq:intGamma}) simply modifies the self-energy
function in (\ref{eq:propag}) as
$\Sigma_a(E,\Gamma)=\Sigma_a(E-i\Gamma/2)$, whence, by
(\ref{eq:FGR}),
\andy{gameffi}
\beq
\gamma_{\rm eff}(\Gamma)=-2\Im
\Sigma_a\left(\omega_a-i\frac{\Gamma}{2}\right)=\frac{4}{\Gamma}
\int d\omega\;
\kappa_a(\omega)
\frac{\frac{\Gamma^2}{4}}{(\omega-\omega_a)^2+\frac{\Gamma^2}{4}}.
 \label{eq:gameffi}
\eeq
The effective decay rate (\ref{eq:gameffi}) is shown in the
central frame of Figure \ref{fig:gammatot} as a function of
$4/\Gamma$. The behavior is similar to that described in Sec.\
\ref{sec-pulsed}. For large values of $\Gamma$ one gets a linear
behavior
\andy{lingameffi}
\beq
\gamma_{\rm eff}(\Gamma)\sim 4/\Gamma \tau_{\rm Z}^2, \quad
\mbox{for}\quad \Gamma\to\infty,
 \label{eq:lingameffi}
\eeq
which, when compared with (\ref{eq:lingammaeff}), yields
Schulman's relation $\tau\simeq 4/\Gamma$ \cite{Schulman98}. When
$\Gamma<\Gamma^*=4/\tau^*$, i.e.\ when the response of the
apparatus is not very quick, the decay is accelerated (IZE). For
$\Gamma \to 0$ one recovers the natural decay rate $\gamma$.

\subsection{Continuous ``Rabi" observation (response time $K^{-1}$)}
\label{sec-Rabi}
\andy{sec-Rabi}
The previous example is nothing but a more refined model of (the
first stage of) a detection process than that given by the
projection prescription. In this sense one might be led to think
that irreversibility is a fundamental requisite for obtaining
quantum Zeno effects: the observed system has to be coupled to a
{\em bona fide} detector that irreversibly records its state. This
expectation would be incorrect. In order to hinder (or
accelerate) decay it is enough to introduce an external agent
which couples differently to the initial state $\ket{a}$ and to
the ``decay" products $(1-\ket{a}\bra{a})\ket{\psi}$ ($\psi$
being the wave function of the system). In other words, one only
needs an interaction which is able to distinguish whether the
system is in its initial state or not: in this (very) loose sense
the external agent can be viewed as a detector
\cite{zenoreview}. Let us illustrate this point by adding to
(\ref{eq:hamtot}) the following interaction Hamiltonian
\beq
H_{\rm meas}(K)=K \left(\ket{b}\bra{M}+\ket{M}\bra{b}\right),
\eeq
which is probably the simplest way to include an external
apparatus: as soon as state $\ket{b}$  becomes populated it
undergoes Rabi oscillations to state $\ket{M}$ with Rabi frequency
$K$ (detector response time $=1/K$) \cite{Panov}. The interaction
modifies the self-energy function as
$\Sigma_a(E,K)=[\Sigma_a(E+K)+\Sigma_a(E-K)]/2$, whence the
effective decay rate reads \cite{induced}
\beq
\gamma_{\rm eff}(K)=\left[\gamma(\omega_a+K)+\gamma(\omega_a-K)\right]/2=\pi
\left[\kappa_a(\omega_a+K)+\kappa_a(\omega_a-K)\right]
\eeq
and is shown in the right frame of Figure \ref{fig:gammatot} as a
function of $1/K$. The behavior is similar to those previously
described. For large values of $K$ one gets the behavior
\andy{lingameffiK}
\beq
\gamma_{\rm eff}(K)\sim \pi \kappa_a(K)\sim \Lambda/\tau_{\rm Z}^2 K^2, \quad
\mbox{for}\quad \Lambda\to\infty.
 \label{eq:lingameffiK}
\eeq
Note, however, that this quadratic law, unlike the linear laws
(\ref{eq:lingammaeff}) and (\ref{eq:lingameffi}), is not generic,
for it depends on the specific asymptotic behavior of the chosen
form factor $\kappa_a$. As in the previous cases, when $K<K^*$,
i.e.\ when the response of the apparatus is not very quick, the
decay is accelerated (IZE) and for $K\to 0$ the system eventually
decays with the natural rate $\gamma$.

\section{Conclusions}
\label{sec-concl}
\andy{sec-concl}

We have shown that the only requisite to obtain QZE is a coupling
which is able to ``pick out" the initial state of the system. For
unstable systems this can also give rise to IZE. The recent
experiment \cite{Raizen01} has proved the existence of a
transition from QZE to IZE in the case of pulsed measurements for
a {\em bona fide} unstable system. It would be interesting to
check the presence of such a transition also in the other cases
envisaged in this paper (continuous and continuous Rabi
observation).

\end{document}